\documentclass[aps,pra,a4paper,reprint,showpacs]{revtex4-1}
\usepackage{amsmath}
\usepackage{amssymb}
\usepackage{graphicx}
\usepackage[utf8]{inputenc}
\usepackage[T1]{fontenc}
\usepackage{ae,aecompl}
\usepackage[english]{babel}
\usepackage{hyperref}

\newlength{\figa}
\graphicspath{{./RFigures/}}

\newcommand{\la}{\left \langle}	
\newcommand{\ra}{\right \rangle}	
\newcommand{\too}{\text{--}}

\date{\today}

\begin{document}
\title{Geometrical defects in two-dimensional melting of
many-particle Yukawa systems}

\author{Ar\={u}nas Radzvilavi\v{c}ius}

\affiliation{Department of Theoretical Physics, Vilnius University, 
Saul\.{e}tekio al.~9-III, LT-10222 Vilnius, Lithuania}
\affiliation{CoMPLEX, University College London,
Gower Street, London W1E 6BT, UK }

\pacs{
52.27.Lw, 
64.60.Cn, 
61.20.-p, 
}

\keywords{complex plasma; two-dimensional melting; polygon construction}

\begin{abstract}
We perform Langevin dynamics simulations and use polygon construction 
method to investigate 
two-dimensional (2D) melting and freezing transitions in many-particle Yukawa 
systems. 2D melting transitions can be characterized as proliferation of 
geometrical defects --- non-triangular polygons, obtained by removing 
unusually long bonds in the triangulation  
of particle positions. A 2D liquid is characterized 
by the temperature-independent number of quadrilaterals and 
linearly increasing number of pentagons. We analyse specific types of 
vertices, classified by the type and distribution of polygons 
surrounding them,  and determine temperature 
dependencies of their concentrations. Critical points in a 
solid-liquid transition are followed by the peaks in the 
abundances of certain types of vertices.
\end{abstract}
\maketitle


\section {Introduction}
For a few decades melting and freezing transitions in two-dimensional 
many-particle systems have been investigated in a variety of experimental 
and computational studies, without reaching a 
definite conclusion regarding its nature. According to the most widely 
accepted
Kosterlitz-Thouless-Halperin-Nelson-Young (KTHNY) theory, melting of a 
single two-dimensional (2D) crystal occurs via two continuous
phase transitions, first from a solid to hexatic phase, and
then from the hexatic to an isotropic fluid \cite{Nelson, Young}. The theory 
also predicts, what it is proliferation of unbound topological defects 
--- dislocations and then free disclinations --- what plays a crucial 
role in 2D phase 
transitions and breaks positional and orientational order. Some 
experiments and theoretical studies, however, suggest a first-order 
grain-boundary-induced melting  
scenario in polycrystalline systems \cite{Nosenko, Chui}.

Although in 2D a true crystalline order 
can not survive at finite temperatures $kT>0$  \cite{Mermin, Hohenberg}, a quasi-long range
translational order is observed at the conditions of strong coupling, 
for example, in complex plasma layers \cite{Knapek} or charged 
colloidal suspensions \cite{Murray}.
Over the years a broad range of 
empirical criteria was developed to accurately determine  
melting and crystallization points \cite{Wang}. Perhaps one of the most 
famous examples is the Lindemann criterion, which has been applied 
extensively in three-dimensional melting and freezing, while its 
generalized version has 
been used in some studies of two-dimensional transitions \cite{Bedanov}. 
Other methods frequently make use of 
topological defect fractions, bond orientational order parameters, 
orientational and positional correlation functions as well as 
Einstein frequencies \cite{Hartman1}. In a recent work, a polygon 
construction method by Glaser and Clark \cite{Glaser1, Glaser2, 
Lansac} was employed to 
characterize transitions in a rapidly heated and cooled two-dimensional 
complex plasma experiment \cite{Suranga}.  

Systems of strongly correlated particles in complex plasmas are of 
particularly high importance in the experimental studies of phase 
transitions. Complex plasma usually consists of polymer microparticles 
immersed  
in a weakly ionized gas, where distinct dust grains are known to 
interact through the Yukawa  
(Debye-H\"uckel) interaction \cite{Konopka}. Convenient time and length scales of 
these systems allow for the direct optical observation  
of collective many-particle phenomena as well as accurate 
measurements of individual particle positions by the means of video 
microscopy \cite{Morfill, Chu, Thomas0}.

In a recent experiment, the phenomenon of superheating was observed in 
a solid state of two-dimensional 
complex plasma \cite{Feng}. It was demonstrated, that  
during a rapid heating process, the concentration of defects can stay low and 
complex plasma can retain the properties of a solid even at the temperatures
above a melting point. The same experimental results were later
analysed by the method of geometrical defects \cite{Suranga}, originally 
developed by Glaser and Clark, which we use in the current work.

Geometrical defects in a polygon construction method are identified by 
removing unusually long bonds in 
the triangulation map of particle positions, so that resulting 
polygons have three or more sides. It was shown, that this method 
provides great sensitivity and unveils some interesting 
features, undetectable by the conventional analysis of topological 
defects \cite{Suranga}. As another measure of disorder, the abundance of
different kinds of vertices, grouped according to the type and order of 
the adjacent polygons, was suggested in the same work. 

In the present contribution we report numerical 
studies of two-dimensional melting and crystallization in 
strongly coupled Yukawa systems. 
Langevin dynamics simulations are performed to simulate gradual
heating and cooling. Critical points are 
determined employing orientational order parameters and topological defect fractions. However, the main motivation behind the current work is to 
present the method of geometrical defects and vertex fractions in the 
polygon construction as a 
sensitive tool to analyse the order-disorder transitions and 
characterize the state of a 2D system. Furthermore, our findings 
resolve the issue of the prominent peaks in the 
temperature-dependencies of certain types of vertex concentrations 
\cite{Suranga}, by showing, that peaks correspond to the critical 
points in the initial and final stages of the order-disorder phase transition. 

In the following section we briefly describe the model system and 
simulation methods, as well as essential tools used in the analysis of 
phase transitions. Main results of the simulations and numerical 
analysis are presented in Section \ref{Results}, while Section 
\ref{Summary} summarizes the article.

\section{Simulation}
\label{Simulation}
\subsection{Model system}
A widely used approximation to describe interactions between 
particles in complex plasmas is the Yukawa inter-particle potential \cite{Konopka}
\begin{equation}
V_{ij}=Q^2 (4\pi\varepsilon_0r_{ij})^{-1} \exp (-r_{ij}/\lambda_{\rm 
D}).
\end{equation}
Here $Q$ is the charge of a particle, $r_{ij}$ is the distance 
between particles $i$ and $j$, $\lambda_{\rm 
D}$ stands for the Debye length, which accounts for the screening of the interaction 
by other plasma species.

Strongly coupled many-particle systems with Yukawa interactions are 
fully characterized
by two dimensionless parameters, namely the coupling strength $\Gamma = 
Q^2/(4 \pi \varepsilon_0 b k_{\rm B}T)$ and 
screening parameter $\kappa = b/\lambda_{\rm D}$, where $b$ is the
two-dimensional Wigner-Seitz radius \cite{Kalman}. In the simulations 
presented here we  
set the screening strength to the constant value of $\kappa=2$.

As a scale of length in our numerical simulations it is convenient to 
choose the Wigner-Seitz radius  
$b$, which is directly related to the areal number concentration of 
particles, $n=1/(\pi 
b^2)$. Therefore, the corresponding scale of energy is $\epsilon = 
Q^2/(4 \pi \varepsilon_0 b)$  
and time is scaled according to the value of an inverse plasma 
frequency
\begin{equation}
\omega_0^{-1}=\left( \frac{Q^2}{ 4 \pi \varepsilon_0 m b^3      
}\right)^{-1/2}.
\end{equation}

The model system consists of $N=2430$ identical particles in a 2D rectangular box of 
area $85.71\times89.07$, interacting via the Yukawa 
potential. Periodic boundary 
conditions are applied and, since the inter-particle potential is 
short-ranged, the cut-off distance is set to $r_{\rm c}=8$. Only 
particle pairs separated by less than $r_{\rm c}$ are taken into 
account in the force calculation.

We study order-disorder transitions in the model system by performing Langevin dynamics 
simulations with slow changes of temperature. Particle  
positions are updated according to the dimensionless Langevin equation
\begin{equation}
\mathbf{ \ddot{r} }_i = -\nabla_i V(\mathbf{r}_1,\dotsc,\mathbf{r}_N)-\mu 
\mathbf{\dot{r}}_i+\mathbf{f}_i,
\end{equation}
where $\mathbf{f}_i$ represents a randomly fluctuating Brownian force. 
In a thermodynamic equilibrium $\left< \mathbf{f}_i(t)\right>=0$, 
while the friction coefficient $\mu$ is related to the  
Gaussian noise $\mathbf{f}_i(t)$ by the
fluctuation-dissipation theorem \cite{Gunsteren}
\begin{equation}
\left< {\rm f}_i(t) {\rm f}_j(t^\prime) 
\right>=2 \mu kT_{\rm ref}\delta_{ij} \delta(t-t^\prime),
\end{equation}
where $i,\, j \in 
\{1,\dotsc,N\}$ and $kT_{\rm ref}$ is the desired target temperature in the 
units of  
$\epsilon$. In our simulations we use $\mu=0.2$. The Langevin equation is 
integrated numerically employing an impulse method of integration \cite{Skeel}.

\subsection{Analysis}

A common way of analysing the structure of 2D many-particle systems is 
calculation of a Delaunay triangulation, which yields a network of 
bonds connecting each particle with its nearest neighbours.
A coordination number can be assigned to each particle, which is a 
number of the triangulation bonds between the particle and its closest 
neighbours. The coordination  
number of a particle in a perfect hexagonal lattice is always equal to 
six. Topological defects are identified as particles with a different 
coordination number,  
usually five or seven.

 Two most common defect types are the disclination 
(a single particle with a non-sixfold coordination) and the dislocation (two 
connected particles with five and  
seven closest neighbours) \cite{Radzvilavicius}. Quite frequently defects 
organize themselves in lengthy chains  
or grain boundaries, indicating a polycrystalline 
structure of the system. As an alternative, a Voronoi construction is sometimes used 
in the context of 2D dusty plasmas, where 
defects are identified as non-six-sided polygons \cite{Nosenko, Feng2, 
Thomas, Feng3}. 

To quantify the abundance of topological defects, we use the
defect fraction $\rm DF$, which is defined as a number of vertices 
with a 
coordination number other than six, normalized to the total number of 
particles $N$ \cite{Feng3}.

The polygon construction is a different way of characterizing defects in 2D 
systems \cite{Glaser1, Glaser2, Suranga} and helps to identify 
empty volumes in 2D liquids \cite{Bernal}. The authors of \cite{Glaser1} analysed 
bond-angle and bond-length probability distributions in dense 2D 
liquids. It was shown, that  
disordered regions of a liquid exhibit multiple peaks in a bond-angle 
probability distribution, with extra peaks corresponding to the square 
lattice. At the 
same time, triangularly-oriented clusters had a single peak and well expressed 
triangular order. The structure of a two-dimensional system therefore was 
described as a square-triangular tiling containing numerous tiling 
faults.

A triangle, which is the only kind of 
polygon in the initial triangulation of particle positions, is 
considered as a non-defective  
entity. To identify geometrical 
defects, certain bonds are removed from the triangulation map, so that 
two polygons sharing a common bond are merged into one. 
Two possible approaches for the selection of bonds were suggested:
either use a bond-length threshold, or identify a bond
that is opposite to the unusually large angle between a pair of
adjoining bonds. In our analysis we follow the authors of 
\cite{Glaser1, Suranga} and use the critical bond-angle of $\alpha=75^ \circ$. 

\begin{figure}[ht]
\centering
\includegraphics[width=0.9\figa]{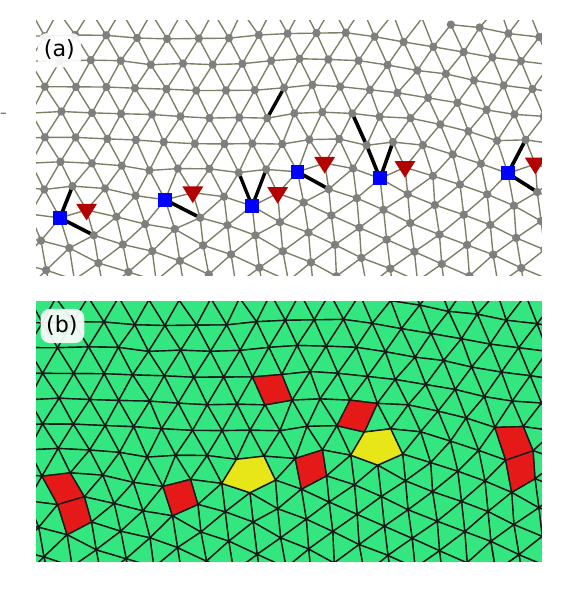}
\caption{Triangulation map of a polycrystalline 2D Yukawa solid, with 
topological defects marked as squares
and triangles (a). By removing the bonds facing unusually large 
angles (marked by bold lines), the polygon construction (b) is 
obtained. In general there is no one-to-one correspondence between 
topological and geometrical defects.}
\label{pc}
\end{figure}

The construction of a polygon map is illustrated in Figure \ref{pc}. 
The first panel (a) shows the 2D triangulation map of particle positions, where most of 
the particles are connected with six closest neighbours. Some vertices, 
however, have five or seven bonds and are marked by small triangles and 
squares. These particles are considered as topological defects and are 
all
the part of a lengthy grain boundary. Triangulation bonds marked by bold 
lines are facing bond-angles larger than the critical value of 
$\alpha=75^ \circ$ and therefore
are selected for removal. The resultant polygon map is presented in 
the second panel (b). Geometrical defects are identified as non-triangular 
polygons, that is, quadrilaterals and  
pentagons. Although the defects appear in a close vicinity of 
the grain boundary, there is no one-to-one correspondence between the polygons 
and topological defects.

The polygon construction contains geometrical information about bond 
lengths and angles, as well as topological information about the 
nearest-neighbour connections. The method has also the advantage of
providing a gradation in the severity of geometrical defects.
Quadrilaterals are the least severe, while pentagons and hexagons
are more severe and indicate large excess volumes. 
Topological defects, on the other hand,  provide only a binary measure 
of local orientational disorder, that is, at the specific location
of a vertex, there either is a defect, or there is
not.

The gradation of defects in the polygon method
allows for a greater sensitivity identifying and classifying
disorder \cite{Suranga}. To characterize the state of our model system 
and the abundance of geometrical defects
we use four distinct order
parameters $P_n=N_n/(2N)$ ($n=3,\, 4,\, 5,\, 6$), defined as the number of polygons with $n$ 
sides $N_n$, normalized to the doubled number of particles $2N$. 

\begin{figure}[ht]
\centering
\includegraphics[width=0.8\figa]{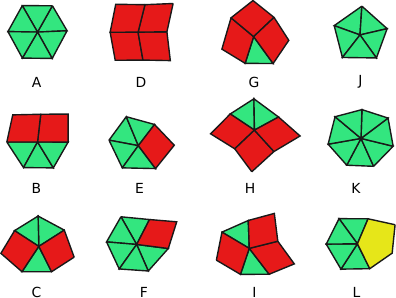}
\caption{Twelve types of vertices, frequently observed in the polygon 
construction of 2D Yukawa systems. Vertices are 
classified by the number and relative distribution of polygons 
around a vertex.}
\label{vertices}
\end{figure}

Figure \ref{vertices} provides a classification scheme for vertices, 
according to the configuration of polygons arranged around them. The 
abundance of different vertex types serves as another way to 
characterize disorder and identify the manner in which polygons cluster together. In 
a perfect crystal, one would observe only vertices of type A. In a regular 
square-triangular tiling, only vertices  
of types A--D are allowed \cite{Glaser1}. Types J and K correspond to the topological 
disclinations, while a vertex L features a severe pentagonal defect.
To quantify the abundance of different vertex types, we calculate
 fractions $X_{\rm Y}$, defined as the number of vertices of a certain
type $\rm Y$ normalized to the total number of particles in the polygon construction.

Unexpected spikes in the time dependencies of parameters $X_{\rm E}$ 
and $X_{\rm F}$ were detected in the analysis of the recent super-heating 
experiment \cite {Suranga}, suggesting that some of the vertices might be 
metastable or exist only in a narrow range of temperatures. One of the 
goals of our work is to resolve this issue.

Phase transitions in two-dimensional systems are usually identified by 
the 
sudden change in orientational or translational order parameters. The local 
orientational order parameter for a particle $j$ is defined as 
\cite{Strandburg}
\begin{equation}
\psi_{6j} = \frac{1}{N_j}\sum_{k=1}^{N_j} {\rm exp} (6i\theta_{jk}), 
\end{equation}
where $\theta_{jk}$ is the angle between the bond connecting particles $j$ and $k$ 
and some fixed direction. $N_j$ is the coordination number of the 
particle $j$. The magnitude $\lvert \psi_{6j} \rvert$ is close to 
unity for a particle inside a hexagonal lattice but is small close to 
the domain walls (grain boundaries) or in a liquid. On the other hand, 
the value of a complex argument $\arg  
(\psi_{6j})$ represents the angular orientation of a 
neighbourhood or entire domain.

The parameter $\psi_6= \lvert \la \psi_{6j} \ra \rvert$, that is, a 
magnitude of the averaged complex orientational order parameter, defines the  
overall orientational order of the system. In 
polycrystalline solids, however, complex numbers $\psi_{6j}$ corresponding to 
the particles from different domains, tend to cancel in the averaging 
process. Therefore, $\psi_6$ approaches very small values in the limit 
of an infinite sample size. 
Another parameter can be used in such cases, namely $\psi_{ \lvert 6 
\rvert}=\la \lvert 
\psi_{6j} \rvert \ra$, which represents the average local orientational 
order of the whole system \cite{Wang}.
%

\section{Results}
\label{Results}

\begin{figure}[ht]
\centering
\includegraphics[width=1\figa]{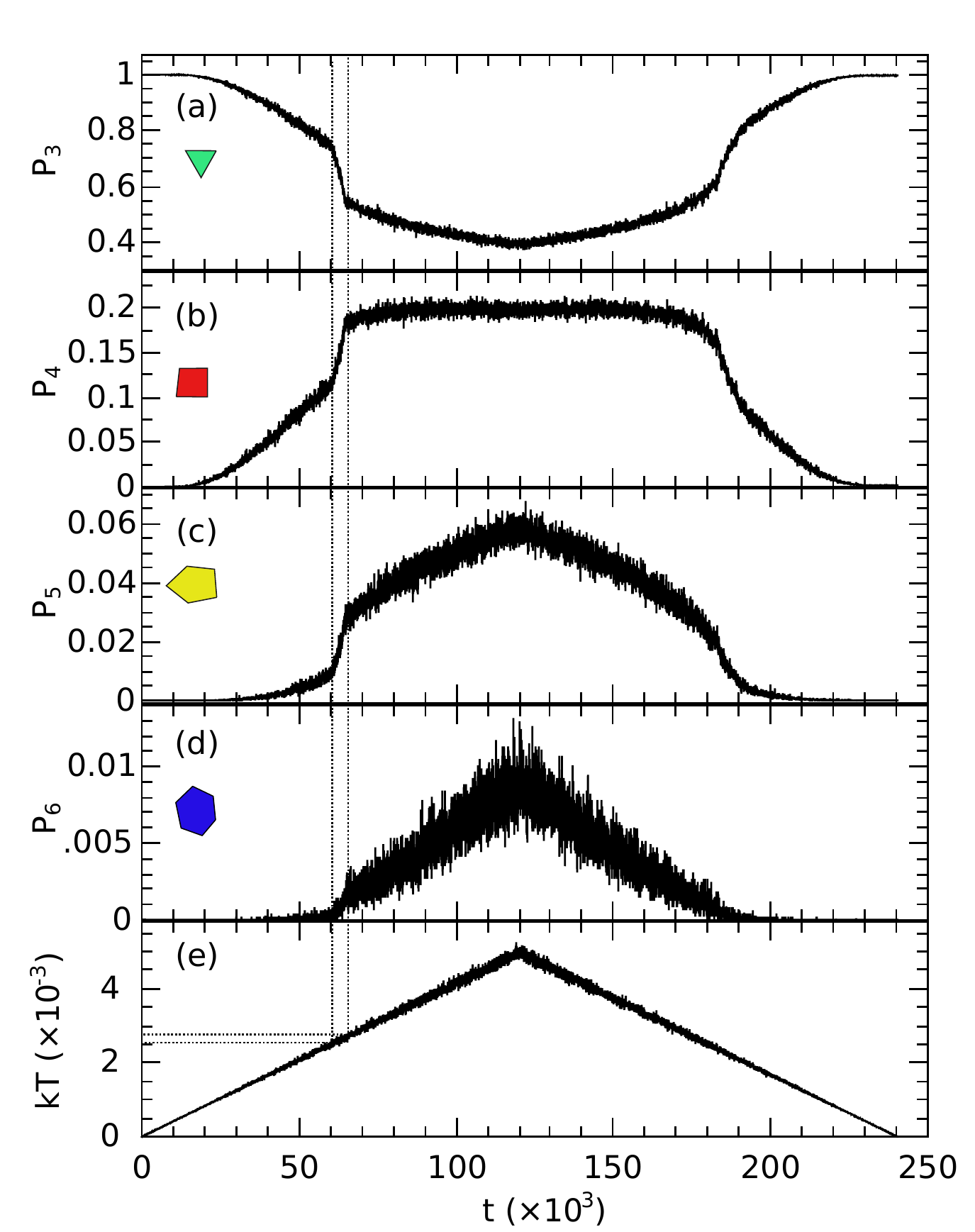}
\caption{Time series for order parameters $P_3$ (a), $P_4$ (b), $P_5$ 
(c) and $P_6$ (d), characterizing the abundances of triangles, 
quadrilaterals, pentagons and hexagons in the polygon construction of 
the 2D 
Yukawa system. Time series of the kinetic temperature $kT$ is 
presented in the last panel (e). Two vertical lines mark the beginning 
and ending points of the melting transition.}
\label{hex-polygons}
\end{figure}

We start our simulations with a defect-free lattice in a strongly-coupled 
state and the temperature of $kT=10^{-7}$. The system is then slowly 
heated over the period of time of $\Delta t=240\,000$, until the temperature
of $kT=0.005$ is reached. 
The stage of steady cooling then follows, 
restoring the temperature to its initial value (see, for example, 
Figure \ref{hex-polygons}-e). The chosen rate of heating is low enough to reach 
the equilibrium at each step of the simulation outside the region of 
a fast order-disorder transition. Therefore, lower rates of heating 
and cooling would not substantially change the qualitative results of our numerical 
experiments, except in the close vicinity of the transition, which we 
discuss later.

The peak temperature is chosen to be well 
above the melting point, as found by the previous studies 
\cite{Hartman1, Hartman2}. The actual kinetic  
temperature of the system is calculated from observed particle 
velocities, $kT=\langle v_i^2 \rangle/2$ and is found to fluctuate 
around the prescribed values of $kT_{\rm ref}$, with the fluctuations 
being proportional to the temperature. We 
keep track of the order parameters and  defect 
fractions throughout the whole cycle. In this section we first 
investigate the 
orientational order of the 2D system, and then turn to the 
defects and polygons.

\begin{figure}[ht]
\centering
\includegraphics[width=1\figa]{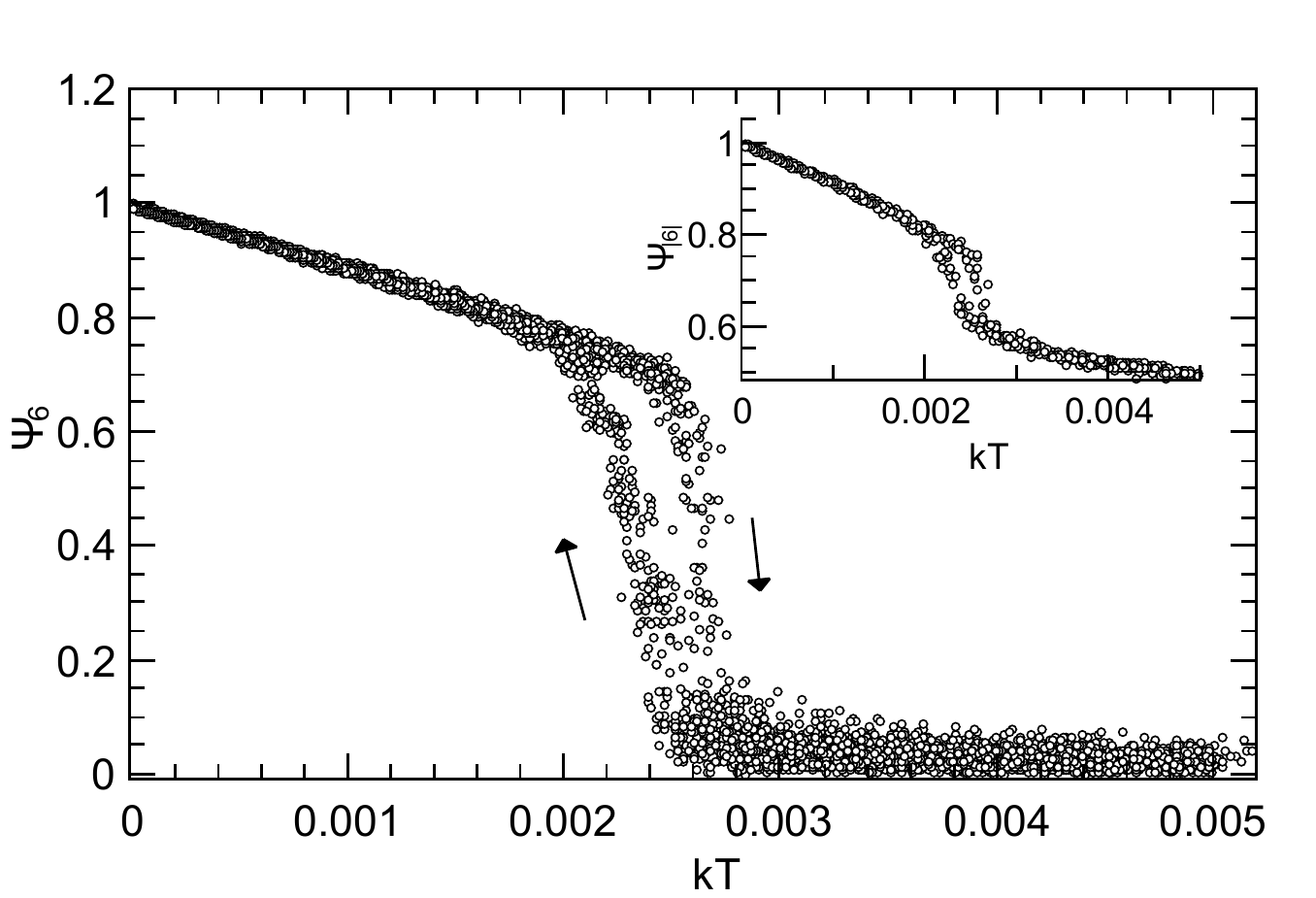}
\caption{Averaged orientational order parameters $\psi_6$ and 
$\psi_{\lvert 6 \rvert }$ of a 2D Yukawa 
system during the heating and cooling simulation cycle.} 
\label{hex-psi}
\end{figure}

The initial values of both orientational order parameters 
$\psi_6$ and $\psi_{ \lvert 6 \rvert}$ are very close  
to unity, as Figure \ref{hex-psi} shows, and correspond to the 
defect-free hexagonal lattice. The system  
exhibits a sudden loss of 
orientational order in the temperature range of $kT=0.0025 \too 
0.0027$ (the corresponding coupling parameter values are $\Gamma=370 \too 400$), which is 
the signature of a melting phase transition. Our 
observations are in a fair agreement with \cite{Hartman1,  
Hartman2}, where phase transitions in two similar 2D systems were 
observed near the values of $\Gamma=384$ and $\Gamma=415$. At the end of the heating 
phase $\psi_6$ drops below $0.1$ and   
$\psi_{\lvert 6 \rvert }$ to the value of approximately $0.5$. 

As the temperature is gradually lowered, a 2D Yukawa liquid freezes 
back to the hexagonal lattice. However, the evolution of the order 
parameters does not follow exactly the  
same path as in the case of melting and hysteresis is observed (Figure 
\ref{hex-psi}).
$\psi_6$ stays low until 
the temperature of  
$kT=0.0024$ is reached, which is lower than the
melting point. Changes in $\psi_{\lvert 6 \rvert }$ also occur at 
somewhat lower 
temperatures and are not as abrupt as in the case of melting. As we 
show later, the 
effect of 
hysteresis is most likely a result of the finite rate of heating and 
cooling. 

\begin{figure}[ht]
\centering
\includegraphics[width=1\figa]{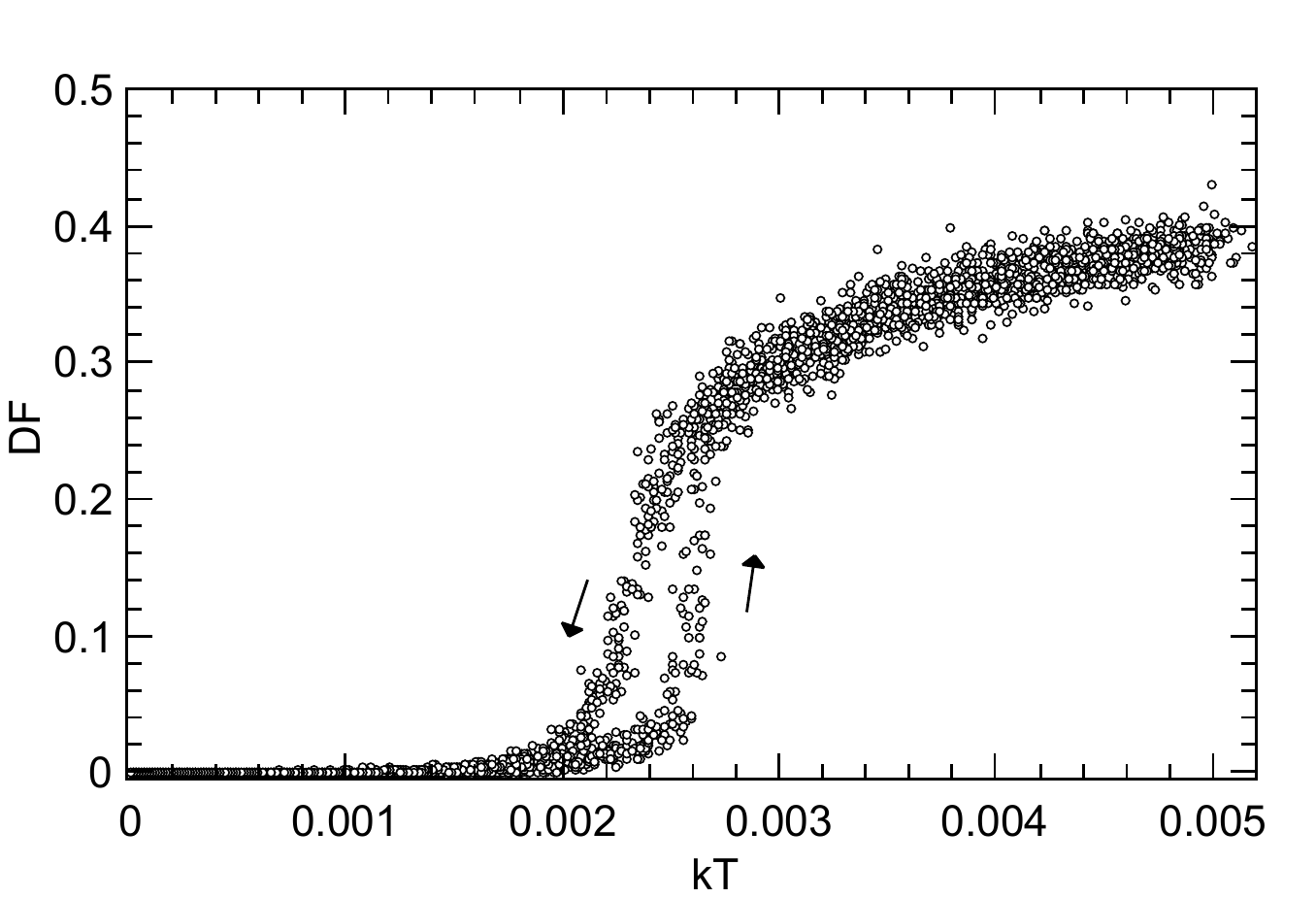}
\caption{Topological defect fraction $\rm DF$ as a function 
of the kinetic temperature $kT$.}
\label{hex-TD}
\end{figure}

The concentration of topological defects during the stage of slow heating exhibits 
a similar trend, with a weak temperature dependence before and after the 
transition, see Figure \ref{hex-TD}. A sudden proliferation of 
defects is observed in the range of $kT\approx 0.0025\too 0.0027$. Changes in the defect 
concentration during the gradual cooling are not as abrupt and occur 
at somewhat lower temperatures, $kT\approx 0.0024\too 0.0022$.

\begin{figure}[ht]
\centering
\includegraphics[width=1\columnwidth]{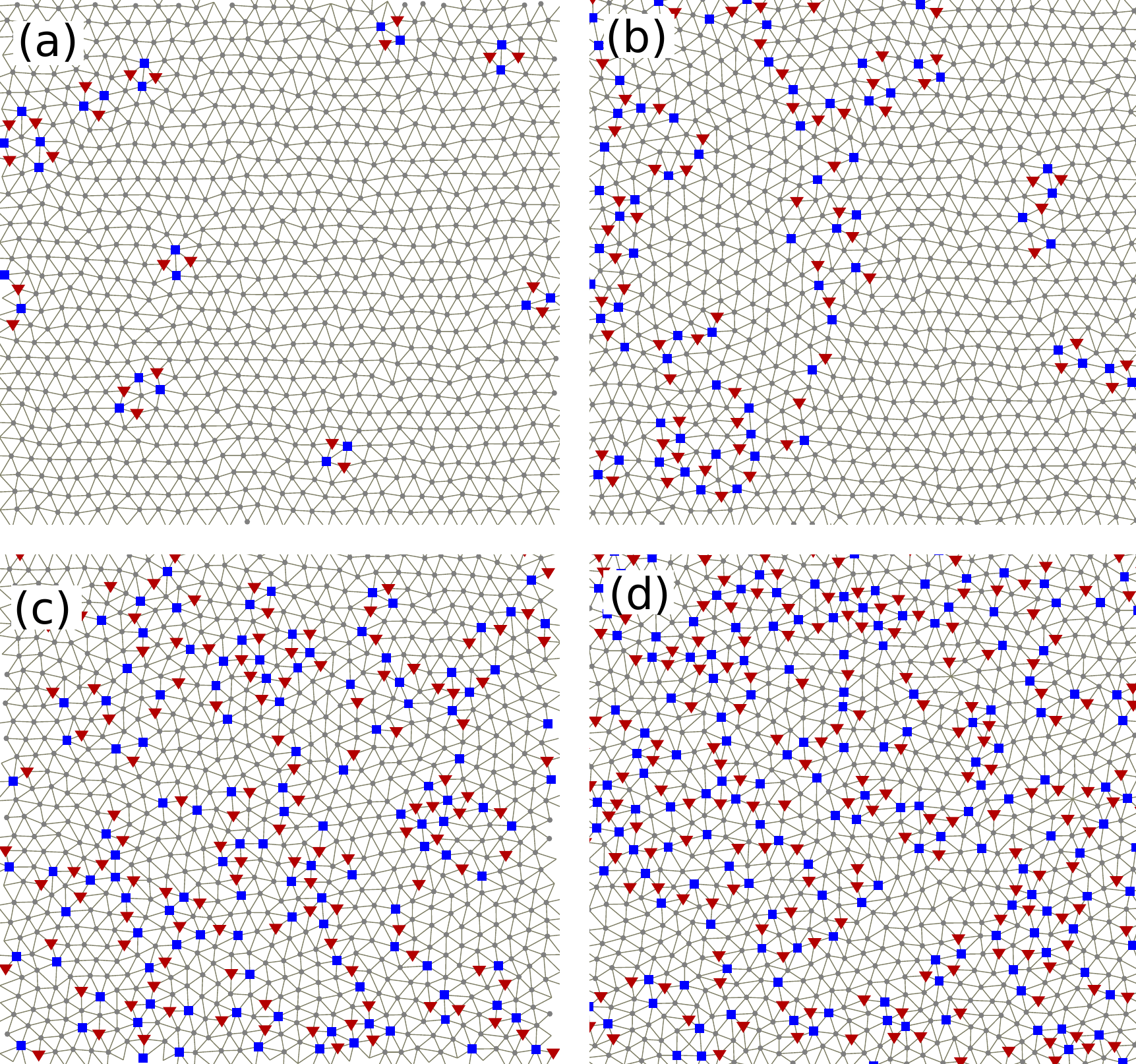}
\caption{Typical arrangements of topological defects during the melting 
transition of a 2D Yukawa system. Triangles here represent particles 
with five closest neighbours, while squares mark 
vertices with the coordination number equal to seven. The corresponding 
temperatures are $kT=0.00245$ (a), $0.0026$ (b), $0.0027$ (c) and 
$0.0045$ (d).}
\label{hex-tdstages}
\end{figure}

Four typical snapshots of particle positions during the 
transition are shown in Figure \ref{hex-tdstages}. Here triangles 
correspond to the particles with only five nearest neighbours, while 
squares mark positions of vertices with seven triangulation bonds. In a solid phase 
(Figure \ref{hex-tdstages}-a), 
defects mostly appear as quartets, composed of two disclinations with five 
bonds and two vertices with seven neighbours. Alternatively, this 
four-defect complex can be treated as a bound pair of two 
dislocations. Apparently, this kind of topological fault does not 
significantly change either positional or orientational order. Larger 
defect complexes, consisting of more than four defective vertices 
emerge before the melting transition. 

As it is illustrated in the second panel of the figure, during the 
melting transition (e.g. $kT=0.0026$) defect complexes grow and 
spread.  However, as it can be seen in Figure \ref{hex-tdstages}-b, 
large defect-free patches still exist, suggesting, that the 
transition from a defect-free lattice to the disordered state is not 
homogeneous. Bound dislocations are still found in the ordered patches, 
however are seldom seen. While some disclinations and 
dislocations are 
present, they are clearly not the main cause of the loss of order; most of 
these defects appear as parts of larger regions of condensed 
defect groups and chains. This observation supports the theory of grain-boundary 
induced  melting \cite{Nosenko, Chui, Quinn}, or possibly the coexistence of hexatic 
and liquid states, as described in \cite{Qi}. Further investigations of larger systems 
would be needed to unambiguously determine the mechanism of the melting 
transition.

Finally, at the end of the transition --- Figure 
\ref{hex-tdstages}-c --- topological defects form 
large inter-connected complexes, destroying orientational order 
completely. Free dislocations and disclinations can still be 
occasionally found in the larger groups of defects.  The distribution 
of defects becomes homogeneous  
in a high-temperature liquid (Figure \ref{hex-tdstages}-d).

As the system is slowly cooled, defect complexes and large defect-free regions can still 
be found at the temperatures as low as $kT=0.0022$, together with 
some free dislocations. At lower temperatures, 
however, these complexes tend to shrink and 
rearrange, eventually leaving only bound and free dislocations as well 
as interstitial particles. The final configuration corresponds to the 
nearly perfect triangular lattice with a few free dislocations, 
leading to the low defect concentration and values of the orientational 
order parameter close to unity.

\begin{figure}[ht]
\centering
\includegraphics[width=1\figa]{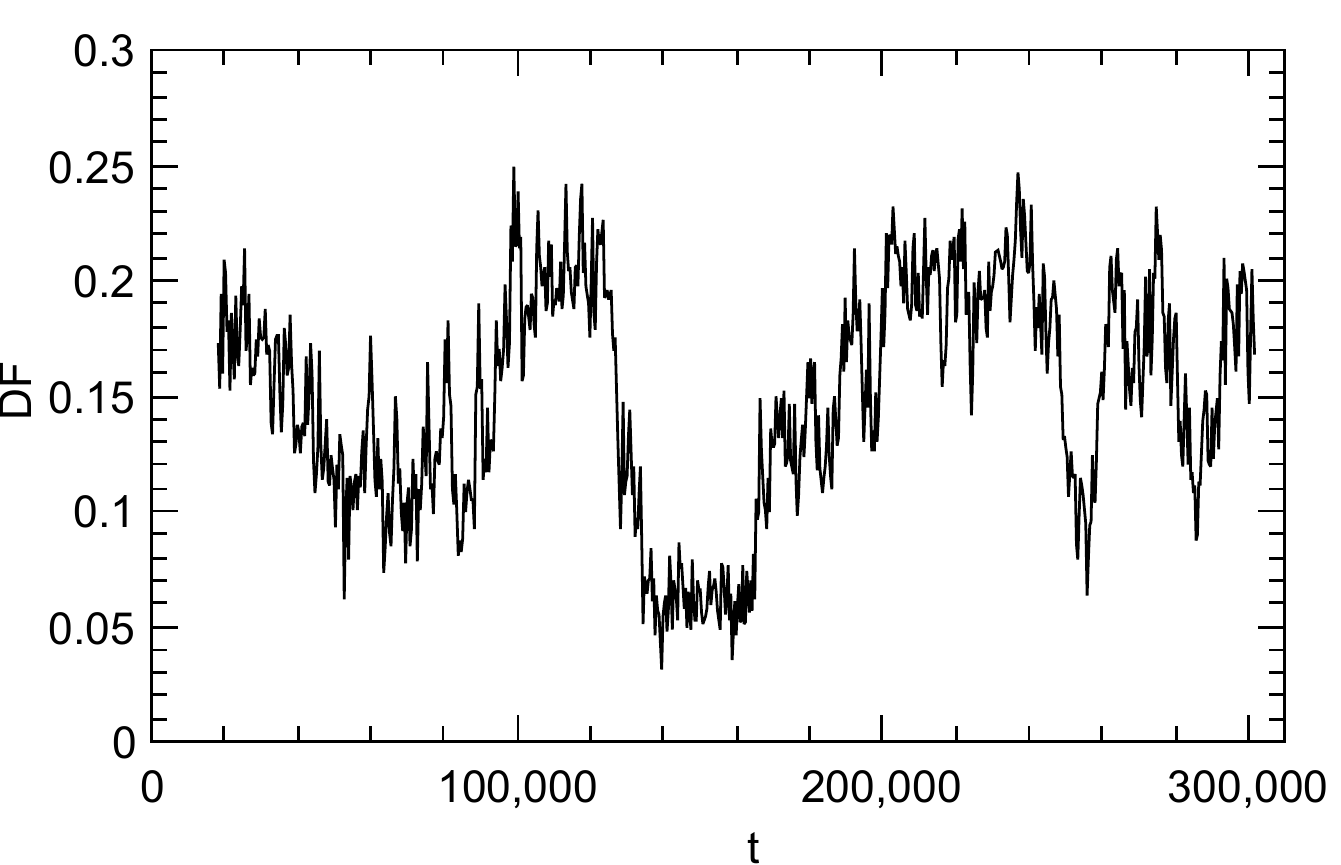}
\caption{Time series for the topological defect fraction $\rm DF$ at 
the prescribed temperature of $kT_{\rm ref}=0.00243$.}
\label{ht}
\end{figure}

Previous studies of similar systems \cite{Schweigert} 
suggest that the effect of hysteresis might be caused by the finite rate 
of heating and cooling. We test this hypothesis by analysing 
the evolution of topological defect fraction at the constant prescribed temperature 
of $kT_{\rm ref}=0.00243$, starting from either liquid or solid initial state. 
According to Figure \ref{ht}, at this temperature the system slowly switches 
between high and low values of the order parameter, 
corresponding to the defect configurations depicted in parts (a) and (c) of Figure \ref{hex-tdstages}. Therefore, there is a 
range of temperatures, in which ordered and disordered states are unstable and have 
approximately the same probabilities to be observed, as reported in 
\cite{Schweigert, Morf}, and no hysteresis should occur in 
the limit of infinite simulation time.  

\begin{figure}[ht]
\centering
\includegraphics[width=1\columnwidth]{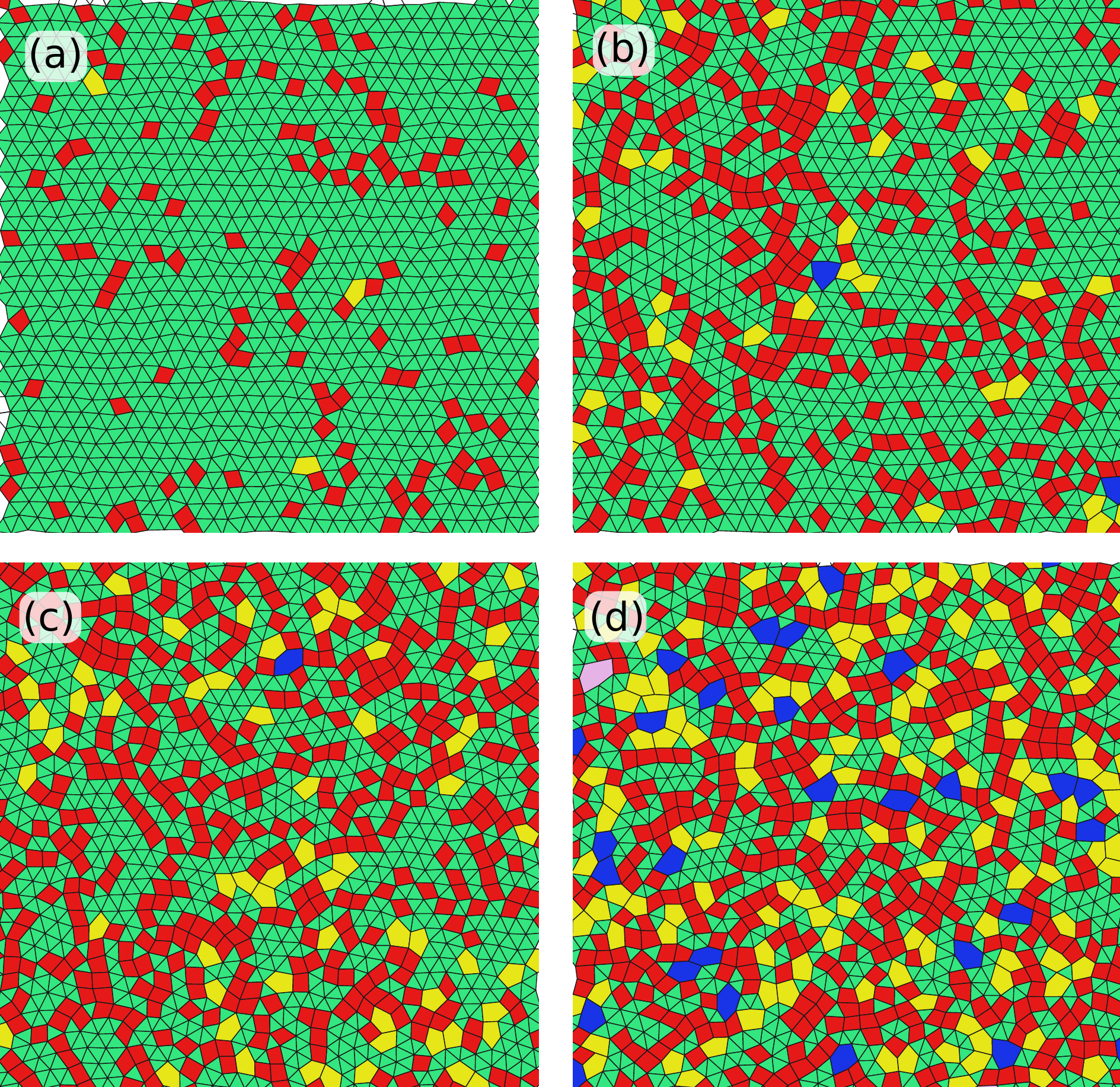}
\caption{Typical arrangements of geometrical defects during the gradual 
heating stage in a
2D Yukawa system: solid phase at $t=45\,002$ (a), configurations during (b) and 
right after (c) the 
solid-liquid transition and high-temperature liquid phase at $t=120\,000$ 
(d). The corresponding temperatures are $kT=0.0018$ (a), $0.0026$ (b), 
$0.0027$ (c) and $0.0045$ (d).}
%
\label{hex-PC}
\end{figure}

Let us now turn to the analysis of the polygon construction. The 
initial defect-free lattice corresponds to 
a triangular tiling. Therefore, the triangle is the only type of 
 polygon present in the original configuration. As the system is 
continuously heated, quadrilaterals and occasional pentagons appear. As it can be seen 
in Figure \ref{hex-PC}, quadrilateral defects tend to cluster 
together, forming long chains or ``ladders'' in a solid phase (a) or larger patches 
of a  distorted square lattice in a liquid (c). Defect-free zones are 
still present during the initial stage of solid-liquid transition, 
e.g. at the temperature of $kT=0.0026$ as depicted in Figure \ref{hex-PC}-b. Solitary hexagons appear much 
later and are most abundant in the high-temperature Yukawa liquid (d).

The quadrilateral is a first type of geometrical defect to 
appear in a low-temperature solid, first seen near the temperature of 
$kT=5 \cdot 10^{-4}$. This is as expected, since the quadrilateral is the 
least severe geometrical defect. Also it is the most abundant 
type of defect in both solid and liquid states. 
The parameter $P_4$, defined as a number of  
quadrilateral defects normalized to $2N$, increases steadily 
as the temperature rises during early stages of heating. As it is 
demonstrated in the second panel of Figure 
\ref{hex-polygons}, the proliferation rate gets significantly higher 
as the temperature reaches the value of $kT = 0.0025$ and the melting 
transition begins. The rapid transition ends at around $kT=0.0027$, 
where the  order parameter fluctuates around the value of
$P_4 =0.20$. 

The abundance of quadrilateral defects in the two-dimensional Yukawa liquid does not change 
significantly as the temperature is further increased. Therefore, we 
suggest, that a 2D  
liquid right after the phase transition can be characterized 
by the temperature-independent value of the quadrilateral order 
parameter close to $P_4=0.20 \pm 0.01$. These observations are illustrated in  
Figure \ref{hex-polygons} as time series for the order parameter (b) and 
temperature $kT$ (e). 

Occasional pentagonal defects are first spotted at the temperature of 
nearly $kT=0.001$. The most significant proliferation of 
these defects takes place in the range of 
$kT=0.0025\too 0.0027$, where the value of a pentagonal order parameter changes 
from $P_5=0.01$ to $P_5=0.03$. After the melting transition, $P_5$ 
increases almost linearly with the temperature, at a constant rate. We 
may conclude, that in the context of 
pentagonal defects, the liquid state can be characterized by values of the
order parameter $P_5>0.03$ and a steady growth in a number of 
pentagons.

In our simulations we observe only a relatively small number of 
hexagonal defects, with the highest value of an order 
parameter close to $P_6 \approx 0.012$. Although the time and temperature dependencies 
of $P_6$ are rather noisy (see Figure \ref{hex-polygons}-d), some 
general observations can still be made. The most noticeable spread of hexagons starts at about 
$kT=0.0025$, which roughly coincides with the start of the
melting transition. Afterwards, $P_6$ seems to increase steadily. 

Geometrical defects arrange themselves around the particles in a 
variety of ways. Some of the most frequent are classified in Figure 
\ref{vertices} as vertex types A to L \cite{Glaser1, Suranga}. In a perfect 
triangular lattice, one would observe only the type A, where six 
triangles join forming a hexagon. In a liquid,  
where quadrilaterals tend to form interconnected complexes and ``ladders'' 
(Figure \ref{hex-PC}), the number 
of vertices B, D, G and H is expected to increase.

As we can see, the 
arrangement of polygons with respect to a certain vertex can be used as 
an indicator of disorder throughout melting or freezing transitions. Therefore, 
we further investigate the evolution of vertex fractions $X_{\rm Y}=N_{\rm Y}/N$, 
defined as the ratio  
of a number of vertices for a certain vertex type $N_{\rm Y}$ to the total number of 
particles $N$.

\begin{figure}[ht]
\centering
\includegraphics[width=1\figa]{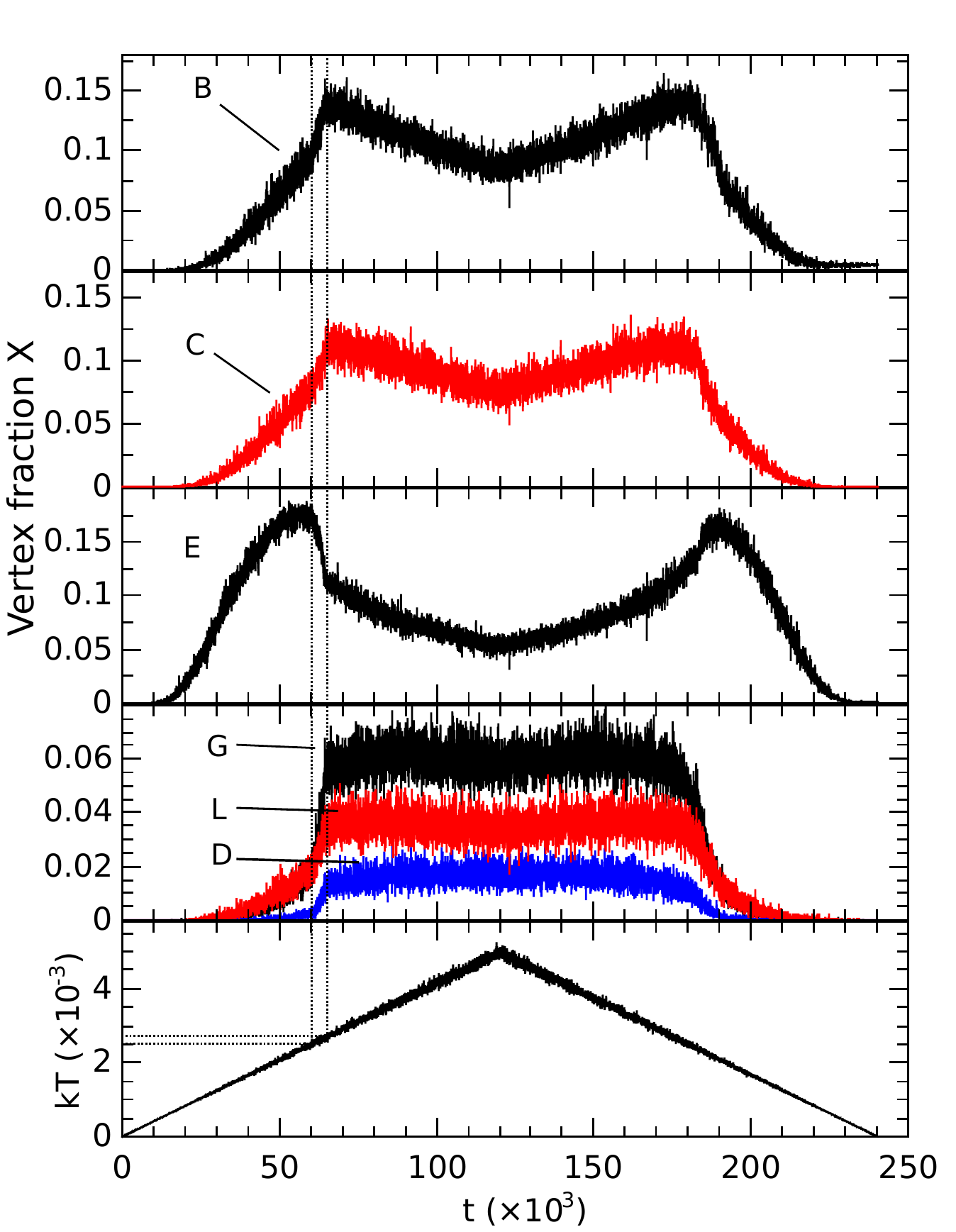}
\caption{Time series for vertex fractions of vertex types B-L and the temperature 
$kT$. The time 
dependence of a vertex fraction $X_{\rm F}$ is virtually identical to the one for 
vertex E and therefore is not shown here. Vertical lines mark the 
beginning and ending points of the rapid order-disorder transition. }
\label{hex-VF}
\end{figure}

Vertices E and F are the first to appear when the 
defect-free system is slowly heated. Just as the quadrilateral defects, 
these vertices are first observed at the  
temperature of $kT \approx 5 \cdot 10^{-4}$. Both vertex types E and F contain a 
single quadrilateral and four or five triangles, and are created by 
removing an inner (E) or outer (F) bond from a vertex type A. Therefore, the
abundances of vertices E and F have essentially 
identical time and temperature dependencies (third panel of Figure 
\ref{hex-VF}).

As the temperature rises, the order parameter $X_{\rm E}$ grows until 
the critical value of $kT\approx 0.0025$ and the fraction 
of $X_{\rm E} \approx 0.175$ is reached. 
As a matter of fact, it is approximately the same temperature that 
marks the start of the rapid
melting transition, i.e. the sudden loss of orientational order, 
the rapid growth in a number of topological 
defects, squares and pentagons. 
As geometrical defects proliferate  
further at higher temperatures, the number of vertices E and F 
monotonically diminishes.  

Vertex types B and C both contain two quadrilaterals and three triangles, 
but differ in the order they are arranged around a vertex. Two 
quadrilaterals of a vertex type B share a common edge (Figure 
\ref{vertices}), while in a vertex C they are separated by a triangle 
or two. 
Vertex concentrations $X_{\rm B}$ and $X_{\rm C}$ depart from zero 
significantly around $kT=0.001$ and grow further with the kinetic temperature. 
The order parameter $X_{\rm B}$ reaches its highest value of $0.14$ at the 
temperature of $kT \approx 0.0027$, which coincides with the final 
transition to the liquid phase (Figure \ref{hex-VF}). Approximately the same is true for the 
vertex fraction of the type C, which reaches 
its highest value $X_{\rm C}\approx 0.12$  close to the temperature of 
$kT \approx 0.0027$. As the 2D liquid is heated 
further, the concentrations of vertices B and C decreases.

Vertex fractions for types D, G, H, L and I all share similar time and 
temperature dependencies (fourth panel of Figure 
\ref{hex-VF}). They all feature a steady growth before the temperature 
of $kT=0.0025$ is reached,  rapid change throughout the melting 
transition and a weak temperature dependence in a liquid state. Right after the 
transition, that is, temperatures above $kT=0.0027$, the vertex fraction D stays close to 
a value of $X_{\rm D}=0.017$, while those for other types fluctuate 
around $X_{\rm G}=0.060$, $X_{\rm H} = 0.035$, $X_{\rm I}= 0.030$, $X_{\rm 
L}=0.039$. 

A few interesting results were obtained in the recent work 
\cite{Suranga}, where the evolution of geometrical defects and vertex 
fractions in a laser-heated complex plasma was 
studied. It should be noted, though, that the goal of the study was to 
investigate solid super-heating and not the temperature dependencies of order 
parameters. The authors did not 
have a chance to vary the temperature of a system gradually and a 
heating source was turned on and off abruptly. 

First, quadrilateral and pentagonal order parameters in a liquid state 
of the experimental system were 
found to be $P_4 \approx 0.19$ and $P_5 \approx 0.05$. These values 
fully agree with and support the results of our simulations.
Secondly, the authors of \cite{Suranga} observed sudden spikes in the 
abundances of vertex types F and E at the times when the heating 
source was 
abruptly turned on and off. Whether it is a 
signature of vertex metastability or a specific temperature dependence  
remained unclear. In the view of our simulation results and, 
specifically Figure \ref{hex-VF}, we conclude that the reason behind the 
spikes is an intrinsic temperature dependence of vertex fractions B, 
C, E and F, 
with spikes corresponding to the temperatures lower than the final 
kinetic temperature of a liquid.

Just as in the case of the orientational parameters and topological 
defect fractions, an evolution of order parameters $P$ and vertex 
fractions $X$ during the slow crystallization does not exactly duplicate the 
behaviour throughout the heating phase. For example, a peak 
in the time dependence of the vertex fraction E corresponds to 
the temperature of $kT \approx 0.0022$, as opposed to the temperature of 
$kT \approx 0.0025$ in the case melting. Critical values of $X_{\rm B}$ 
and $X_{\rm C}$ are also slightly shifted to lower 
temperatures. In a final configuration after the crystallization, 
there are seven quadrilateral defects and 
a corresponding small fraction of vertices B and E (Figure \ref{hex-VF}).

We have repeated our simulations starting with polycrystalline 
configurations, obtained from a rapidly cooled 2D Yukawa liquid. 
The evolution of order parameters throughout the initial heating phase 
turned out to be slightly configuration-dependent. For example, in a 
few cases orientational order parameters increased as the 
polycrystalline solid was heated, while at the same time 
topological defects diminished. This can be explained by the melting 
of grain boundaries and simultaneous merging of crystalline domains.
Nevertheless, the point of a final transition to the liquid phase was 
found to be the same as in the  
case of the hexagonal initial configuration, that is, $kT \approx 
0.0027$. 
Therefore, most of the results presented here would also hold for the melting 
of a polycrystalline solid.

\section{Summary}
\label{Summary}

In this contribution we report the results of Langevin dynamics simulations, 
performed to investigate two-dimensional melting and 
crystallization transitions in many particle Yukawa systems, such as those
found in complex plasma experiments. To characterize the state of a 
system, we use a polygon construction method, in which  
unusually long  
bonds are deleted from the triangulation map of particle positions. 
Geometrical defects are identified as non-triangular polygons, while 
vertices are classified according to the type and order of polygons 
surrounding them. We also make use of a topological 
defect fraction and orientational order parameters as 
conventional tools to analyse phase transitions.

 
In our simulations of the system with the constant value of screening 
strength ($\kappa=2$), the 
solid-to-liquid 
phase transition takes place in the coupling parameter 
range of $\Gamma \approx 400\too 370$, where rapid changes in order 
parameters and defect fractions are detected. The orientational and 
translational order is destroyed 
by the non-homogeneous growth of large defect complexes and chains, contrary to the KTHNY 
theory of two dimensional melting, which predicts the emergence of 
isolated dislocations and disclinations.

To quantify the disorder, we use polygonal order parameters $P$, that 
is, the normalized number of 
geometrical defects of a certain kind. It turns out, that a liquid 
phase can be characterized by the  
temperature-independent value of a quadrilateral order parameter  of 
$P_4 \approx 0.20$. In the context of  
pentagonal defects, the liquid state is characterized by the value of 
$P_5>0.03$. 

Concentrations of vertices containing three or more 
quadrilaterals (D, G, H and I) show only a weak dependence on the 
temperature in the liquid state, showing the tendency for 
quadrilaterals to cluster together. Temperature dependencies of vertex 
fractions of the types B, C, E and F all feature well expressed peaks 
at the beginning (E, F) or final stages (B, 
C) of the solid-liquid transition with a very similar behaviour during 
the recrystallization.

\section*{Acknowledgements}

Computational resources were provided by Vilnius University, Institute 
of Theoretical Physics and Astronomy.

\end{document}